\documentclass[article,twocolumn,authoryear]{elsarticle}

\usepackage{amssymb}
\usepackage{graphicx}
\usepackage{txfonts}

\usepackage{natbib}

\newcommand{\fig}[1]{Fig.~\ref{#1}}
\newcommand{\eqref}[1]{Eq.~(\ref{#1})}
\newcommand{\dd}{{\rm d}}
\newcommand{\ac}[1]{#1}
\journal{Icarus}

\begin{document}


\onecolumn{\title{Runaway gas accretion and gap opening versus type~I migration}

\author[label1,label2]{A. Crida}
\ead{crida@oca.eu}
\author[label3]{B. Bitsch}
\address[label1]{Laboratoire Lagrange (UMR7293), Universit\'e C\^ote d'Azur / Observatoire de la C\^ote d'Azur, Boulevard de l’Observatoire, CS 34229, 06300 Nice, \textsc{France}}
\address[label2]{Institut Universitaire de France, 103 Boulevard Saint-Michel, 75005 Paris, \textsc{France}}
\address[label3]{Lund Observatory, Department of Astronomy and Theoretical Physics, Lund University, Box 43, 22100 Lund, \textsc{Sweden}}
}

\begin{abstract}
Growing planets interact with their natal protoplanetary
  disc, which exerts a torque onto them allowing them to migrate in
  the disc. Small mass planets do not affect the gas profile and
  migrate in the fast type-I migration. Although type-I migration can
  be directed outwards for planets smaller than $20-30 M_\oplus$ in
  some regions of the disc, planets above this mass should be lost
  into the central star long before the disc disperses. Massive
  planets push away material from their orbit and open a gap. They
  subsequently migrate in the slower, type~II migration, which could
  save them from migrating all the way to the star. Hence, growing
  giant planets can be saved \ac{if and} only if they can reach the gap opening
  mass\ac{, because this extends their migration time-scale, allowing them to eventually survive at large orbits until the disc itself disperses.}

  However, most of the previous studies only  measured the torques on planets with fixed masses and orbits to determine the migration rate. Additionally, the transition between type-I and type-II migration itself is not well studied, especially when taking the growth mechanism of rapid gas accretion from the surrounding disc into account. Here we use isothermal 2D disc simulations with FARGO-2D1D to study the migration behaviour of gas accreting protoplanets in discs. We find that migrating giant planets always open gaps in the disc. We further show \ac{analytically and numerically} that in the runaway gas accretion regime, the growth time-scale is comparable to the type-I migration time-scale, indicating that growing planets will reach gap opening masses before migrating all the way to the central star in type-I migration \ac{if the disc is not extremely viscous and/or thick. An accretion rate limited to the radial gas flow in the disc, in contrast, is not fast enough. When gas accretion by the planet} is taken into account, the gap opening process is accelerated because the planet accretes \ac{material originating from its} horseshoe region. This \ac{allows an accreting planet to transition to type-II migration before being lost} even if gas fails to be provided for a rapid enough growth \ac{and the gap opening mass is not reached}.
\end{abstract}

\date{Accepted\,: October 11, 2016}



\maketitle


\twocolumn 

\section{Introduction}

Planets form in proto-planetary discs. For the terrestrial planets of
the solar system, only Moon to Mars sized embryos need to be formed in
the proto-solar nebula\,; the final assembly of Venus and the Earth is
thought to take place after the gas dispersal, through giant impacts
among these embryos and leftover planetesimals
\citep{Kleine-etal-2009,Jacobson-etal-2014,Raymond-etal-2014}. In
contrast, giant planets must have acquired their final mass while gas
was still present, as they are mostly composed of this gas. In the
core-accretion model \citep{Pollack-etal-1996}, the gaseous envelope
is slowly accreted around a solid core of $\sim 10$ Earth
masses. Hence, gas giants must experience planet-disc interactions
from the mass of an embryo all the way to that of Jupiter. A result of
these interactions is planetary migration, which modifies the orbital
radius of a planet, and generally moves it closer to the central star.

Giant planets open a gap around their orbit, separating the disc
between an inner and an outer disc \citep{LinPapaloizou1986a}. In
typical proto-planetary discs, this happens for planets more massive
than Saturn or Jupiter \citep{Crida-etal-2006}. Once a gap is open,
the planet should be locked between the inner and outer disc, and
follow the viscous evolution of the proto-planetary disc, that is in
general a slow accretion towards the central star
\ac{\citep{LinPapaloizou1986b,Nelson-etal-2000}}. This is called
type~II migration, and could explain the semi-major axis distribution
of giant exoplanets \citep[e.g.][]{Lin-etal-1996}, especially when
photo-evaporation of the protoplanetary disc is taken into account
\citep{Alexander-Pascucci-2012}. It should be noted that the above
description is ideal, and in reality a planet can decouple more or
less from the disc evolution
\citep{Crida-Morby-2007,Duermann-Kley-2015}, and some gas can pass
through the gap \citep{LubowDAngelo2006}. Despite these recent
developments, the global picture that \ac{single} gap opening planets
migrate slowly and roughly \emph{together with} the disc still holds.

In contrast, smaller planets --\,which do not perturb significantly
the density profile of the disc\,-- migrate \emph{with respect to} the
disc, in the so-called type~I migration regime \citep{Ward1997}. In
this regime, the migration rate is proportional to the planet
mass. Hence, type~I migration \ac{is not a big issue for the embryos of the terrestrial planets, but} has always been an issue for growing
giant planets\,: accreting a gaseous envelope should take way longer
for their solid cores than migrating all the way into their host star
(the typical migration time-scale for a $30$ Earth mass body is only
$10\,000$ orbits). In fact, the first planet population synthesis
models \citep{Alibert-etal-2005,Ida-Lin-2008,Benz-etal-2008,Mordasini-etal-2009}
had to decrease the efficiency of type~I migration by a factor 100 at
least, if they wanted planets to survive.

In the past decade, huge progress has been made on type~I migration,
mainly relative to the corotation torque \citep[see][for a complete
  review]{Baruteau-etal-PPVI}. It has been shown that this torque can
be positive, and overcome the classical, negative, differential
Lindblad torque when the disc is not isothermal
\citep{Paardekooper-Mellema-2006,Kley-Crida-2008,BaruteauMasset2008AD,Kley-etal-2009}. Typically,
this is efficient for planets in the $5-30 M_\oplus$ range, in the
inner regions of the disc, where the radial gradient of entropy is
steep (ex\,: at opacity transitions). An analytical formula for the
torque felt by a planet in type~I migration has been found by
\citet{Paardekooper-etal-2011} based on 2D numerical simulation, and
confirmed in 3D radiative numerical simulations by
\citet{Bitsch-Kley-2011} and more recently by \citet{Lega-etal-2015}
who included stellar irradiation. Combined with an accurate
description of the temperature and density profiles of the
protoplanetary disc and its evolution, this allows to produce
migration maps, where the torque felt by a planet is given as a
function of its mass and position in the disc
\citep{Bitsch-etal-2013_I,Bitsch-etal-2014_II,
  Bitsch-etal-2014_III,Baillie-etal-2016}. In such maps, it appears
possible to block a \ac{planet} at a zero-torque radius, where it can
grow slowly.

\ac{However, above a critical mass $M_{\rm crit}$, the corotation
  torque always saturates and vanishes. In general $M_{\rm
    crit}\approx 20 M_\oplus$, depending on the opacity, density and
  viscosity of the disc. Therefore, the too fast type~I migration
  problem is solved only below $M_{\rm crit}$. As $M_{\rm crit}$ is 5
  to 10 times smaller than the gap opening mass, the question of the
  fast inwards migration of giant planets, as they grow from $M_{\rm
    crit}$ until they open a gap, remains open. In this paper, we
  address this critical question (and only this question). We study in
  which conditions a growing giant planet can open a gap before type~I
  migration drives it all the way into its host star. }

\ac{After having presented our set-up, code and units in
  section~\ref{sec:setup}, we first study briefly the opening of a gap
  by a \emph{migrating} giant planet in section~\ref{sec:gap}. This
  study is necessary because \citet{Malik-etal-2015} recently
  suggested} that giant planets would not be able to open their gap if
they are migrating too fast. More precisely, they argue in favour of
\citet{Hourigan-Ward-1984} who stated that if the planet crosses its
corotation region faster than a gap opens, the gap never opens and the
planet \ac{therefore remains in type~I migration. If this is true,
  there is no hope for a giant planet to ever open a gap when its
  grows past $M_{\rm crit}$. In this case, the survival of giant
  planets would become a puzzle because they should never leave the
  type~I migration regime. We show in contrast that a Jupiter mass
  planet does open a gap in about a hundred orbits, even if it
  migrates fast. }

\ac{Second, we compare the growth and migration time-scales in
  section~\ref{sec:acc}. We show analytically and with numerical
  simulations that once a $20\,M_\oplus$ core starts its runaway gas
  accretion, it should reach the gap opening mass before its
  semi-major axis is dramatically reduced. Finally, one may worry that
  the theoretical runaway accretion rate can not be sustained if the
  disk can not provide gas fast enough. However, we show in
  section~\ref{sec:suck} that if this occurs, then the planet has
  already opened a gap because all the gas in its horseshoe region has
  spread to the accretion streams. Consequently, once the runaway
  accretion of gas starts, nothing prevents the opening of a gap by
  the growing giant planet. After a discussion in
  section~\ref{sec:discuss}, we summarize our findings in
  section~\ref{sec:conclu}.}

\section{Units and simulations set-up}
\label{sec:setup}

\ac{In this paper, all our simulations are performed with the
  FARGO-2D1D code \citep{Crida-etal-2007}, which is a 2D grid code in
  polar ($r$-$\theta$) coordinates. The resolution is always
  $dr/r=0.01=d\theta$, unless specified otherwise. Far from the
  planet, the standard 2D grid is replaced by a 1D grid, assuming
  azimuthal symmetry, allowing to model the viscous spreading of the
  disc. This is crucial for an accurate modelization of type~II
  migration, but not necessary for a study of type~I migration. Even
  if our focus here is not type~II migration, but the transition into
  type~II migration, we choose to use this code given the negligible
  computing cost of the 1D grid.}

\ac{Our units are $L$ as the arbitrary length unit (generally the initial
  orbital radius of the planet), the mass of the central star $M_*$ as
  the mass unit, and we set the gravitational constant $G=1$ so that
  an orbital period at $L$ is given by $P_L=2\pi T$ with
  $T=\sqrt{L^3/GM_*}$ the time unit.}

\ac{The equation of state is locally isothermal, because we focus on
  cases where the thermal part of the corotation torque would be
  saturated. The aspect ratio is always uniform (no flaring) with the
  usual value $h=H/r=0.05$, so that the sound speed is given by
  $c_s=\frac{0.05}{\sqrt{r/L}}\,(L/T)$. Unless stated otherwise, the
  gas viscosity is given by \citet{Shakura-Sunyaev-1973}'s
  prescription, with a rather standard value of $\alpha=10^{-3}$. With
  this setting, a Jupiter mass planet (that is\,: a planet whose mass
  is $M_J=10^{-3}M_*$, or mass ratio to the star is $q=10^{-3}$) has a
  gap opening parameter as defined by \citet{Crida-etal-2006} of
  $2/3<1$, so we expect this planet to open a gap. The threshold for
  gap opening according to this criterion (that is\,: the gas density
  in the middle of the gap should be $10\%$ of the unperturbed gas
  density) is $0.436\,M_J$.}

\ac{The initial surface density of the gas disc is always of the form
  $\Sigma(r)=\Sigma_0\times (L/r)$. With such a slope of the surface density
  profile and an $\alpha$-prescription for the viscosity in a
  non-flared disc, the viscous torque exerted by the disc inside any
  radius $r$ on the disc outside $r$ is independent of $r$. Hence, the
  viscous torque on any elementary ring is zero, and the disc is at
  equilibrium, with no radial viscous drift of the gas. Of course, at
  the inner (resp. outer) edge of the disc, there is no support from
  an inner (resp outer) disc and the gas will spread. This
  perturbation to the density profile will propagate inside the disc
  profile at the viscous rate. Nonetheless, for most of the times we
  will consider here, the disc around the planetary orbit should not
  spread, and type~II migration is expected to be very slow.}

\ac{The semi-major axis of the planet is noted $a$, and its orbital
  angular velocity is $\Omega=\sqrt{GM_*/a^3}$. The gravitational
  potential of the planet is smoothed using the usual so-called
  $\epsilon$-smoothing, with $\epsilon=0.6\,r_H$ where
  $r_H=(q/3)^{1/3}a$ is the Hill radius of the planet. The
  self-gravity of the gas disc is not taken into account. Whenever the
  torque exerted by the disc on the planet is computed, the region
  within $0.6\,r_H$ is excluded, using a smooth Fermi function, as
  prescribed by \citet{Crida-etal-2009_CPD}. }

\section{Gap opening by migrating giant planets}
\label{sec:gap}

\subsection{Fixed mass with forced migration}

\begin{figure}
\includegraphics[width=\linewidth]{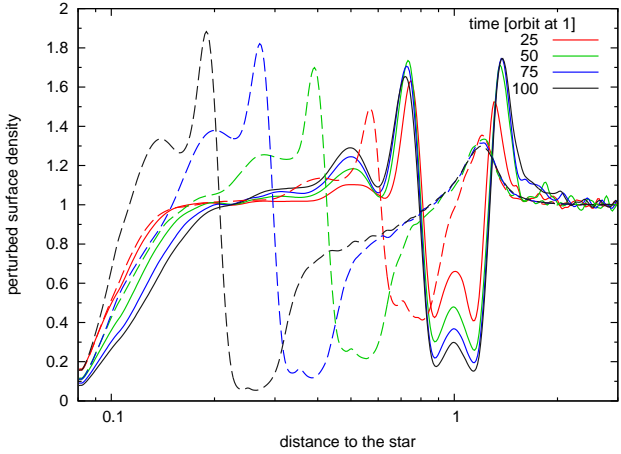}
\caption{Perturbed surface density profiles
  $\Sigma(r)/\Sigma_0(r)$. Different \ac{colours} correspond to
  different times, labelled in \ac{$P_L$, the orbital period at
    $r=1$. Solid curves correspond to the case of a non migrating
    Jupiter, while dashed curves show the case of a migrating Jupiter
    with constant $\tau_m = |a/\dot{a}| = 72\,P_L$ such that the
    planet migrates from $a_0=L$ to $a_f=0.25\,L$ in $100\,P_L$.}  The
  $x$-axis extends between \ac{0.08 (inner edge of the 1D grid) and 3
    (where the outer edge of the 2D grid is at 2.5)}, with a
  logarithmic scale for convenience because the gap width is
  proportional to the orbital radius.  }
\label{fig:Jup}
\end{figure}

\ac{To start with, we have performed simulations in which a Jupiter
  mass planet is thrown\footnote{More precisely, the mass of the
    planet is smoothly increased from $0$ to its final value $M_J$
    over the first period $P_L$. It remains constant afterwards.} into
  an unperturbed disc of initial density $\Sigma_0(r) = 10^{-4}
  (L/r)\ M_*/L^2$. The 2D-grid spans the radial range $0.1-2.5$ and
  the 1D-grid extends from $0.08$ to $10$. The planet starts at
  $a_0=1$, and is forced to migrate at a constant rate
  $\tau_m=-a/\dot{a}$ in such a way that it reaches $a_f=0.25$ in
  $100\,P_L$. The resulting perturbations to the density profile (that
  is\,: $\Sigma(r)/\Sigma_0(r)$\,) are shown in \fig{fig:Jup} after
  25, 50, 75, and 100 orbits, as the dashed curves. The solid curves
  show for comparison the density profiles at the same times in the
  case where the planet remains at $a=1$. As one can see, the gap
  opening is not impeached by the migration, although the planet
  crosses the width of the gap every 25 orbits. The gap is actually
  deeper in the migrating planet case, because 100 orbits at $r=L$
  correspond to more orbits at $r=0.25\,L$. Simulations with shorter
  migration times (namely $50$ and $25\,P_L$) show the exact same
  thing\,: the gap opens at the same pace, whether the planet migrates
  or not. We conclude that the gap opening process takes about a
  hundred orbits to complete, and the planet may migrate during this
  time, but the gap opening itself is not affected by the migration.}

\ac{Note that at the inner edge of the 1D-grid, the open boundary
  condition allows the gas to flow inwards, and the density
  drops. This drop propagates outwards at the slow, viscous rate, but
  does not perturb our simulations\,: after $100\,P_L$, the density is
  still the unperturbed one at $0.25\,L$ if the planet does not
  migrate.}

\ac{The migrating planet seems to act as a snowplough, pushing gas
  inside its orbit and leaving a slightly depleted outer disc
  behind. Although some gas can cross the planetary orbit (otherwise
  the outer disc would be empty from the planetary orbit up to $r=1$),
  this behaviour is expected as the planet carries its depleted
  horseshoe region during its migration (see below) and repels the
  inner disc with its gravitational torque. If the planet was let
  free to migrate, it would be slowed down or blocked by the overdense
  inner disc, and transition into type~II migration. Actually, after
  its forced migration during $100\,P_L$, the planet is released free
  and we see it migrate slightly outward.  }

\subsection{Forced growth with free migration}
\label{sub:fgfm}

\ac{To go further, we have performed simulations in which
the planet grows from $M_0=3\times 10^{-5}\,M_*$ (or $10.0\,M_\oplus$
if $M_*=M_\odot$) following\,:
\begin{equation}
M_p = M_0 \exp(t/\tau_g)\ .
\label{eq:expgrowth}
\end{equation}
We use the same code and parameters, except for a lower $\alpha=4\times
10^{-4}$ (in order to slow down the viscous evolution of the disc),
where the 1D grid extends from 0.02 to 5 and 2D grid from 0.3 to
2.3\,.}

\begin{figure}
\includegraphics[width=\linewidth]{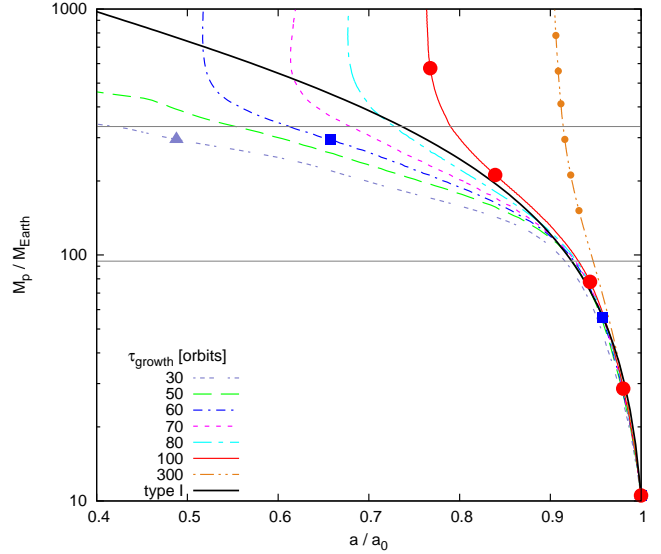}
\caption{Tracks of planets in the Mass$-$semi-major-axis plane,
  following Eq~(\ref{eq:expgrowth}), with various, constant values of
  $\tau_g$ and \ac{always} $\tau_m/\tau_g=100(M_0/M_p)$ \ac{whatever $\tau_g$}. The thick, plain black
  line is the track corresponding to pure type~I migration,
  \eqref{eq:track}. The thin horizontal lines mark
  $94.4\,M_\oplus$, the gap-opening mass and $333\,M_\oplus =
  M_J$. Symbols are placed every 100 orbits on the colour lines
  corresponding to $\tau_g=30$, $60$, $100$, and $300$ (only after
  $800$ orbits in this latter case)\,; owing to the logarithmic scale
  of the $y$-axis, the symbols are evenly spaced vertically.}
\label{fig:R100}
\end{figure}

The planet starts at $a_0=L$ and is left free to migrate. In the
type~I regime, the planet feels a torque $\Gamma = -A\Gamma_0$ where
$\Gamma_0=(M_p/M_*)^2\Sigma\, a^4\Omega^2h^{-2}$ and $A$ is a constant
numerical coefficient given for a locally isothermal EOS by
$A=2.5+1.7\beta_T-0.1\alpha_\Sigma-1.1\left(\frac32-\alpha_\Sigma\right)$,
with $\alpha_\Sigma$ and $\beta_T$ the negatives of the power indices of
the density and temperature profiles, respectively
\citep{Paardekooper-etal-2010}. The migration rate is then given by\,:
\begin{equation}
\dot{a} =
-2A\left(\frac{M_p}{M_*}\right)\frac{\Sigma\,{a}^3\,\Omega}{M_*\ h^{2}}
 = -\frac{a}{\tau_m}\ .
\label{eq:typeI}
\end{equation}
As well known, $\tau_m\propto \left(\Sigma M_p\sqrt{a}\right)^{-1}$, while here
$\tau_g$ is constant. Hence, assuming that $a$ does not vary much, we
can set $\Sigma_0(L)$ such that the ratio of the growth and migration
rates follows $\tau_m/\tau_g = 100(M_0/M_p)$ during the
simulation. Then, combining the expressions of $M_p$ and $\dot{a}$
leads to\,:
\begin{equation}
M_p=M_0\times\left[1-100\ln(a/a_0)\right]\ .
\label{eq:track}
\end{equation}
This is shown as the thick, solid black curve in \fig{fig:R100}. The
planets should follow this curve as long as they are in type~I
migration. \ac{Admitedly, with $\tau_m/\tau_g = 100(M_0/M_p)$, the
  migration rate is smaller than the growth rate at low masses, and
  the planets are expected to cross only the Hill radius of a Jupiter
  mass planet by the time they reach the gap-opening mass. However, 3
  Hill radii of a Jupiter mass planets are then crossed by the time
  Jupiter's mass is reached. The key here is that the time it takes to
  this point is set by $\tau_g$ and $\Sigma_0$, so that the migration
  can be arbitrarily fast in this critical phase\,; this allows to
  test the gap opening by a fast migrating planet. A smaller number
  than $100$ in \eqref{eq:track} would result in a shallower black
  line in \fig{fig:R100}. Because we want for this experiment the gap
  opening mass to be reached before the planet has migrated into its
  host star, this ratio of the migration and growth times seems
  appropriate.}

Depending on the value of $\tau_g$, \ac{the planets} should evolve
more or less slowly along \ac{the thick solid black curve}, but as
long as they follow Eqs~(\ref{eq:expgrowth}) and (\ref{eq:typeI}) they
must be on this curve. One expects the planets to depart from this
track progressively as they open a gap, that is in the next $\sim 100$
orbits after they reach the gap opening mass (which is \ac{here}
$94.4\,M_\oplus$ according to the \citet{Crida-etal-2006}
criterion). In particular, \ac{as mentionned in Section~\ref{sec:setup},}
the disc is stationary, and a planet in an ideal type~II migration
regime should have $\dot{a}=0$ and follow a vertical line in the $M-a$
diagram of \fig{fig:R100}. Some moderate migration rate beyond the gap
opening mass could be observed (see discussion about type~II migration
in the introduction), but still much slower than the type~I migration
rate.

The dashed, coloured curves correspond to different $\tau_g$ (and
different $\Sigma_0$ to keep $\tau_m/\tau_g = 100(M_0/M_p)$). To
illustrate the different speeds at which the curves are followed,
symbols have been placed every $100\,P_L$ on a few lines (see
caption)\,; in particular, $300\,M_\oplus$ are reached after $100$,
$200$, $1000$ $P_L$ in the cases $\tau_g=30,\ 60,\ 300$
respectively. The case $\tau_g=100$ (plain red line with large
bullets) illustrates very well the expected behaviour\,: type~I
migration up to $\sim 100\,M_\oplus$, and then transition to type~II
migration about 100 orbits later. In the case $\tau_g=300$, the gap
opening is faster than the growth of the planet, so that the planet
starts carving a partial gap even before reaching $100\,M_\oplus$ and
leaves the type~I track to enter smoothly a type~II like regime. In
the cases where $\tau_g<100$, the planet appears to accelerate with
respect to the expected type~I migration rate as soon as it starts
opening a gap. This is because of the positive feedback caused by the
coorbital mass deficit (cmd)
\citep{Masset-Papaloizou-2003}. Considering that for giant planets,
the horseshoe width is the Hill radius, we can define the maximum
possible cmd as\,:
\begin{equation}
{\rm cmd}_{\rm max} = 2\pi a\Sigma_0(a)\times 2r_{\rm Hill}
= 0.87\,\Sigma_0(a)\,a^2\left(\frac{M_p}{M_J}\right)^{1/3}
\label{eq:cmdmax}
\end{equation}
The ${\rm cmd}$ is proportional to $\Sigma_0$ (hence to $1/\tau_g$),
and if it is larger than the planet mass, it leads to the runaway,
type~III migration
\ac{\citep{Masset-Papaloizou-2003,Peplinski-etal-2008-II}}. Despite
the positive feedback provided by the cmd in massive discs, all
planets eventually open a deep gap and enter type~II migration beyond
Jupiter's mass, even if they have migrated by more than the width of
the gap by the time they reach this mass.

Only in the cases of very massive discs ($\tau_g\leqslant 50$) the
planets hit the inner edge of our 2D grid before they have a chance to
transition to type~II migration. It should be noted that less than a
hundred orbits were spent after the planets reached $94.4\,M_\oplus$,
so they did not have time to open a clean gap. They may also have
entered the type~III migration regime\,: in the case $\tau_g=30$,
$\Sigma_0(L)=7.25\times 10^{-4}\,M_*/L^2$, so that the ${\rm cmd}_{\rm
  max}$ is $4.2 \times 10^{-4}\,M_* = 140\,M_\oplus$ for a
$100\,M_\oplus$ mass planet at $a=L$. We believe that this is the kind
of phenomenon that \citet{Malik-etal-2015} have seen, as already
observed by \citet{Crida-2009}. Actually, if the planet mass is below
the inertial mass limit $q_{\rm limit}$ given by
\citet{Ward-Hourigan-1989}, then $q<0.42\,h^{9/4}\,({\rm cmd}_{\rm
  max}/M_*)^{3/4}$, which is about an order of magnitude smaller than
${\rm cmd}_{\rm max}$ for $h < 0.15$ and ${\rm cmd}_{\rm
  max}/M_*>10^{-5}$. Thus, a strong positive feedback from the cmd,
possibly leading to type~III migration, is easier to reach than the
inertial mass limit\footnote{In order to have $q_{\rm limit}=10^{-3}$,
  such a high gas density is needed that the disc would be
  gravitationally unstable\,: $\pi\Sigma r^2=M_*$ at the location of
  the planet for $\alpha_\Sigma=1$ and $h=0.1045$.}.

\ac{To summarize, because of the positive feedback on migration caused
  by the coorbital mass deficit, a planet opening a partial gap can
  migrate fast on very long distances in dense discs. However, this
  does not prevent the planet from opening a gap eventualy. Either this
  planet is actually not massive enough to open a deep gap anyway, or
  it should open a deep gap but this takes about a hundred orbits\,;
  during this time, the planet may migrate far (and leave the
  simulation frame), but this is only a transcient phase until the gap
  opens and the planet transitions to type~II migration, as
  illustrated by the blue curve labelled $60$ in \fig{fig:R100}.
  There is no such thing as crossing one's horseshoe width faster than
  a gap opens\,: a migrating planet does carry its horseshoe region
  with it. This is known to give rise to the dynamical corotation
  torque \citep{Paardekooper-2014,Pierens-2015}. When the planet and
  the disc have a relative radial motion, the horseshoe has a
  trapezoidal shape in the $r-\theta$ frame moving with the planet,
  but it still exists (see Fig.~8 of \citet{Masset-2008} or Fig.~1 of
  \citet{Masset-Papaloizou-2003}\,). Therefore, when a migrating
  planet starts depleting its horseshoe, it carries a depleted
  horseshoe region, and keeps depleting it, until it becomes empty and
  a gap opens, as shown by \fig{fig:Jup}. In massive discs, as a
  planet starts opening a gap, the coorbital mass deficit accelerates
  significantly its type~I migration (which is already fast), possibly
  leading to the runaway, type~III migration. Admitedly, the planet
  may then migrate far and be lost in its host star before it has time
  to open a gap and transition to type~II migration. But we want to
  stress here that this is not because type~I migration prevents the
  gap opening, but because migration is faster than the gap opening.}

\ac{We have shown that a planet above the gap opening mass should eventually open its gap, even if it migrates fast during the gap opening process. The question now is\,: can a planet reach the gap opening mass before being lost by type~I migration when it's above $M_{\rm crit}$~? In this section, we have assumed arbitrary growth rates. In the next section we relax this assumption and use gas accretion rates determined by previous high resolution 3D hydrodynamical simulations \citep{Machida-etal-2010}, and compare them with migration rates.
}

\section{Runaway gas accretion versus type~I migration}
\label{sec:acc}

In the standard core accretion model \citep{Pollack-etal-1996}, once a
solid core of $\sim 10 M_\oplus$ is formed (phase 1), it accretes
slowly a gaseous envelope (phase 2). When the mass of the envelope
exceeds that of the core, the envelope becomes unstable and collapses
onto the core\,; more gas comes in and the planet undergoes runaway
gas accretion (phase 3). Phase 2 is limited by the cooling rate to
evacuate the energy of the accretion of solids, and could last
millions of years. In this case, $\tau_g \gg \tau_m$ and the planet
migrates almost at constant mass. A planet in this regime would just
go wherever type~I migration drives it, eventually being trapped at a
zero-torque radius. In this section we assume that the core has
outgrown phase 2 and starts its runaway gas accretion in phase 3. At
this point the planet becomes massive enough so that its corotation
torque saturates. During this growth and migration phase, the planet
\ac{generally} loses a major part of its semi-major axis \ac{in planet
  population sythesis studies,} making this part of its evolution
crucial for the final orbital position of the planet at disc dispersal
\citep{Bitsch-etal-2015b}.

\ac{\subsection{Analytical study}}

\citet{Machida-etal-2010} found an accretion rate in the runaway
regime\,:
\begin{equation}
\dot{M}_{p,M10} = \Sigma H^2 \Omega\times\min\{0.14\,;\, 0.83(r_H/H)^{9/2}\}\ ,
\label{eq:MdotMachida}
\end{equation}
where $r_H$ is the Hill radius of the planet, $H$ is the disc scale
height, and the index $M10$ denotes \citet{Machida-etal-2010}'s rate.

\ac{Let us compute the ratio of the typical times for the growth of $M_p$ and the migration\,: $\mathcal{G} \equiv \frac{\dot{a}/a}{\dot{M}_{p,M10}/M_*}$. Using \eqref{eq:typeI}, and \eqref{eq:MdotMachida} one finds\,:
\begin{equation}
\mathcal{G} = -\frac{A\,M_p}{h^4\,M_*}\frac{1}{\min\{0.14\,;\, 0.83(r_H/H)^{9/2}\}}\,.
\label{eq:G}
\end{equation}
This ratio is independent of $\Sigma$, $a$, and
$\Omega$. $\mathcal{G}$ is only a function of $h$ (that we have
assumed uniform here) and $M_p$. For reference, with $h=0.05$ and
$A=2.73$, $\tau_g = \tau_m$ for $M_p= 0.4\,M_J$. Planets smaller than
that grow faster in runaway regime than they should migrate in type~I
migration, hence they should reach the gap opening mass ($0.436\,M_J$
with $h=0.05$ and $\alpha=10^{-3}$) before being lost. }

\begin{figure}
\includegraphics[width=\linewidth]{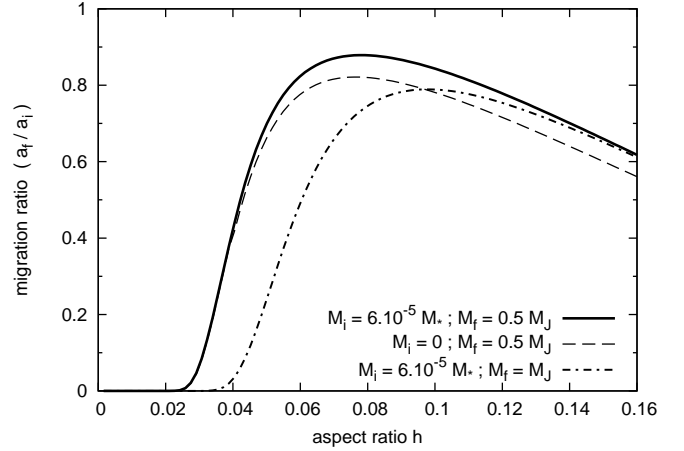}
\caption{\ac{Ratio of the initial to final semi major axis of a planet
    that migrates in pure type~I migration following \eqref{eq:typeI}
    (with $A=2.5$) while growing from an initial mass $M_i$ to a final
    mass $M_f$ in the runaway regime following
    \eqref{eq:MdotMachida}.}}
\label{fig:analytic}
\end{figure}

\ac{Assuming a planet is in type~I migration, following \eqref{eq:typeI} and assuming it grows in the runaway mode following \eqref{eq:MdotMachida}, one can compute analytically the reduction of its semi-major axis while it grows from a mass $M_i$ to a mass $M_f$\,:
$$
\ln\left(\frac{a_f}{a_i}\right)  = \int_{M_i}^{M_f} \dd\ln a = \int_{M_i}^{M_f} \mathcal{G} \frac{\dd M_p}{M_*}\ .$$
Note $M_t=3\left(\frac{0.14}{0.83}\right)^{2/3}\,h^3\,M_*$ the mass at
which the transition occurs in \eqref{eq:MdotMachida} --\,that is\,:
for $M_p>M_t$, the $\min$ term in \eqref{eq:MdotMachida} is $0.14$ and
for $M_p<M_t$ the second argument of the $\min$ should be
considered. For $h<0.176$, $M_t<0.5\,M_J$, so in reasonable discs, the
transition will occur before half a Jupiter mass is reached, and we
will consider hereafter that $M_t<M_f$. For $h<0.0403$,
$M_t<20\,M_\oplus$, so in thin discs it is possible that $M_t<M_i$. In
this case, the above integration is straightforward and leads to\,:
\begin{equation}
\ln\left(\frac{a_f}{a_i}\right) =  -\frac{A}{0.28\,h^4}\left[\left(\frac{M_f}{M_*}\right)^2-\left(\frac{M_i}{M_*}\right)^2\right]
\label{eq:ratioeasy}
\end{equation}
If $M_f>M_t>M_i$, the integral should be split and in the end\,:
\begin{eqnarray}
\hspace{-0.9cm}\ln\left(\frac{a_f}{a_i}\right)\! & = &\! -\underbrace{\frac92\left(\frac{0.14^{1/3}}{0.83^{4/3}}\right)}_{\sim3.0}\,A\,h^2\left\{\left[\left(\frac{M_f}{M_t}\right)^{2}\!-1\right]+4\left[1\!-\!\sqrt{\frac{M_i}{M_t}}\right]\right\} \label{eq:ratioall}\\
 & < & -3\,A\,h^2\left[\left(\frac{M_f}{M_t}\right)^{2}+3\right]\label{eq:ratioallmax}
\end{eqnarray}
Taking $A=2.5$, $M_i=20\,M_\oplus$ and $M_f=0.5\,M_J$ or $M_f=M_J$, the ratio
$a_f/a_i$ provided by \eqref{eq:ratioall} for $h>0.0403$ and
\eqref{eq:ratioeasy} for $h<0.0403$ is plotted as a function of $h$ in
\fig{fig:analytic}. The case $M_i=0$ actually corresponds to
\eqref{eq:ratioallmax} and is shown as the thin, long-dashed curve\,; it shows
that the influence of $M_i$ is actually small, compared to the
influence of $M_f$. However, the real key parameter appears to be
$h$\,: in thin discs ($h<0.035$), migration is faster than growth and
the planets are lost before they reach half a Jupiter mass. In thicker
discs ($h>0.045$), $a_f/a_i>0.6$ and standard type~I migration should
not make the planets migrate by more than $40\%$ of their initial
semi-major axis by the time they reach half a Jupiter mass.
}

\begin{figure}
\includegraphics[width=\linewidth]{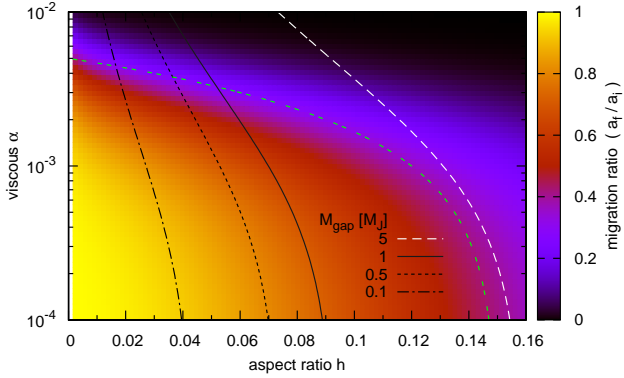}
\caption{\ac{Colour map\,: ratio of the initial to final semi major
    axis of a planet that migrates in type~I migration following
    \eqref{eq:typeI} while growing from $0$ to the gap opening mass in
    the runaway regime following \eqref{eq:MdotMachida}, or slower or
    faster rates but such that $A =
    (\Gamma/\Gamma_0)\,/\,(\dot{M}_p/\dot{M}_{p,M10}) = 2.5$. Black
    and white lines\,: contours of the gap opening mass $M_{\rm gap}$
    as given by \eqref{eq:Mgap}. Green double-dashed line\,:
    $\alpha=0.005-h/30$ (crude proxy for the $a_f/a_i=0.4$ contour
    line). These calculations do not depend on the surface density of
    the gas disc $\Sigma$, but neglect the positive feedback on type~I
    migration caused by the coorbital mass deficit\,; this phenomenon
    is important when a planet opens a partial gap in a dense disc. }
}
\label{fig:migaplane}
\end{figure}

\ac{As we are interested in reaching the gap opening mass, it is
  tempting to replace $M_f$ by the gap opening mass $M_{\rm gap}$,
  which is a function of $h$, as given by Eq.~(10) of
  \citet{Baruteau-etal-PPVI}\,:
\begin{equation}
\frac{M_{\rm gap}}{M_*} = \frac{100}{\mathcal{R}}\left[(X+1)^{1/3}-(X-1)^{1/3}\right]^{-3}\ ,
\label{eq:Mgap}
\end{equation}
  with $\mathcal{R}=a^2\Omega/\nu = 1/(\alpha\,h^2)$ and
  $X=\sqrt{1+3\mathcal{R}h^3/800}$. Substituting $M_f$ in
  \eqref{eq:ratioallmax} by $M_{\rm gap}$ given by \eqref{eq:Mgap}, we
  can compute an upper boundary of $a_f/a_i$ as a function of $h$ and
  $\alpha$, independent of $M_i$\,\footnote{Note that in the limit
    $h\to 0$, the right member of \eqref{eq:ratioallmax} converges
    towards $-3A\frac{\alpha}{10^{-4}}$\,, but does not
    diverge.}. \fig{fig:migaplane} shows this result as a colour map\,;
  it shows that for low enough $h$ and $\alpha$, a planet reaches the
  gap opening mass before its semi major axis would be halved by pure
  type~I migration. In particular, the green double-dashed curve
  corresponds to $30\alpha + h = 0.15$ and is a good proxy of the
  $a_f/a_i=0.4$ contour line. For any fixed $h$, the growth rate is
  determined\,; then, as $\alpha$ increases, $M_{\rm gap}$ increases
  so that migration has more time to reduce the semi-major axis. The
  black and white curves overlaid on the colour map show contours of
  $M_{\rm gap}$ as given by \eqref{eq:Mgap}. For $\alpha=10^{-2}$,
  whatever $h$, type~I migration can divide the semi major axis by
  more than an order of magnitude by the time $M_{\rm gap}$ is reached
  in the runaway growth regime. Conversely, for a fixed $\alpha$, the
  increase of $h$ leads to a larger gap opening mass but also a more
  favourable growth to migration times ratio\,; as a consequence, even
  in thick discs ($h>0.1$) where $M_{\rm gap}> M_J$, planets in
  runaway gas accretion can theoretically reach the gap opening mass
  before being lost into their host star, even if their migration is
  more important in thicker discs. In standard discs ($h\sim 0.05$,
  $\alpha\sim{\rm a\ few}\times 10^{-3}$), $M_{\rm gap}$ is between
  Saturn and Jupiter's masses, and is reached in the runaway growth
  while the orbital radius is roughly halved by type~I migration. }

\ac{This study, being purely analytic, relies on 2 simplified
  hypothesis\,: (i) the planets migrate in pure type~I migration, as
  given by \eqref{eq:typeI}\,; (ii) they grow in the runaway regime
  according to \eqref{eq:MdotMachida}. It allows to explore the
  parameter space in the $h-\alpha$ plane, and to understand the role
  of these two parameters --\,which are the only free ones in the
  final equation, with $A$, which can be seen as the ratio of the
  migration and growth rates normalized to \eqref{eq:typeI} and
  \eqref{eq:MdotMachida}\,:
  $A=(\Gamma/\Gamma_O)\,/\,(\dot{M}_p/\dot{M}_{p,M10})$. Namely,
  $\alpha$ is only responsible for gap opening, while $h$ determines
  the growth and migration rates as well as gap opening. However, we
  have seen in section~\ref{sub:fgfm} that in dense discs, when
  planets start opening a partial gap, they leave the pure type~I
  regime, and they can migrate faster. Consequently, in next
  subsection, we explore with numerical simulations the role of the
  surface density of the disc, which had disappeared from the
  equations here, and explore the effect of a decreased accretion
  rate.}

\subsection{Numerical study}

\ac{To study numerically the competition between migration, runaway
  gas accretion and gap opening, we} perform simulations in which the
increase of the planet mass \ac{follows \eqref{eq:MdotMachida}. As for
  the value of $\Sigma$ in \eqref{eq:MdotMachida}, we take} the
initial density of the disc at the planet location, \ac{$\Sigma_0(a)$}
because in \citet{Machida-etal-2010}'s shearing box, the density
profile can not be perturbed by the planet (no gap
opening). \ac{Furthermore, we are mostly interested in the phase
  before the gap opening, where the density close to the planet is
  actually little perturbed\,; we come back to this choice later.}
Note that here, we prescribe this growth rate to the planet mass,
without removing the gas from the disc. \ac{We set the initial mass of
  the planet to $6\times 10^{-5}M_*=20\,M_\oplus$, which is roughly
  the mass at which planets in the outer disc start their runaway gas
  accretion phase in \citet{Bitsch-etal-2015b} \citep[see
    also][]{Lambrechts-etal-2014}. In these} simulations, the 1D grid
extends from $0.2$ to $50$ and the 2D grid from $1.0$ to $25.126$, and
the planet starts at $10$. \ac{As usual, the disc parameters are
  $h=0.05$ and $\alpha=10^{-3}$, resulting in $M_{\rm gap}=0.436\,M_J$,
  and $a_f/a_i\gtrsim 0.72$ according to \fig{fig:migaplane},
  independent of $\Sigma_0$.}

\begin{figure}
\includegraphics[width=\linewidth]{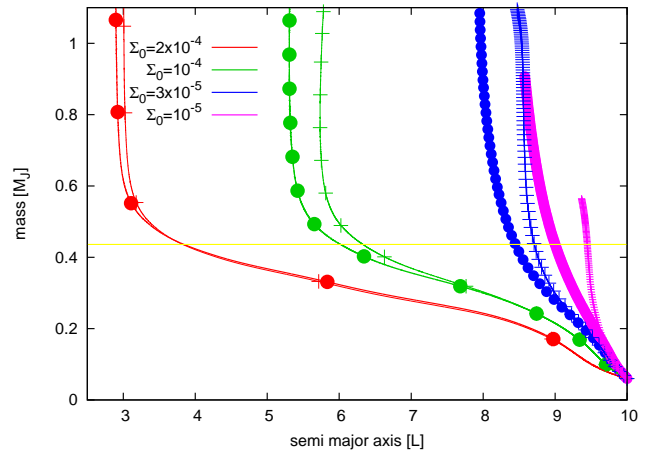}
\caption{Tracks of planets in the Mass$-$semi-major-axis plane,
  accreting following \eqref{eq:MdotMachida} (thick curves with
  dots), or following \citet{Kley1999}'s recipe limited by
  \eqref{eq:MdotMachida} (thin curves with $+$ symbols, \ac{see
    section~\ref{sec:suck}}). The planets are free to migrate in discs
    with locally isothermal equation of state and different surface
    densities. The $\bullet$ \ac{and $+$} symbols are placed on the
    curves every $1000\,P_L$ to show the migration and accretion
    rates. The grey thin horizontal line marks $0.436\,M_J$, the
    threshold mass for gap opening by gravity. }
\label{fig:M}
\end{figure}

The results \ac{of these simulations} are shown in Fig.~\ref{fig:M}
\ac{in the $M_p-a$ plane} as the thick curves, with dots every
\ac{$1000\,P_L$}. Each curve corresponds to a different value of
$\Sigma_0$, given by the key. \ac{All our planets reach the gap
  opening mass} before being lost into their central star.


Concerning migration, all planets in our simulations eventually open a
gap and \ac{almost} stop migrating, as expected for ideal type~II
migration in our disc. \ac{In the case $\Sigma_0=3\times 10^{-5}$, the
  final semi major axis is $\sim 8$, in agreement with the analytic
  estimate above.} We also recover the same trend as previously\,: at
a given mass, planets in \ac{denser} discs migrate slightly faster
because of the positive feedback exerted by the coorbital mass deficit
of the horseshoe region, hence they end further inwards. \ac{Note that
  with $\Sigma_0=10^{-4}$, ${\rm cmd}_{\rm max}=4\,M_J$ at $a=10$ for
  a $0.1\,M_J$ planet. This was not taken into account in the
  analytical calculations of previous subsection, nor was considered
  the time needed to open the gap and transition from the type~I to
  the type~II regime.}

\subsection{Influence of the accretion rate}
\ac{Gas accretion can influence slightly the migration rate of giant
  planets \citep[e.g.][]{Nelson-etal-2000,Peplinski-etal-2008-I}. In
  the frame of our study, the main effect of the accretion rate is on
  the growth of the planet. Admittedly, \citet{Machida-etal-2010}'s
  accretion rate is probably an upper estimate\,: their simulations
  considered a shearing box with constant density boundaries, and did
  not take the gap opening into account (which only plays a role after
  a gap is opened, though). Therefore, taking always
  $\Sigma_0=10^{-4}$, we have divided the accretion rate given by
  \eqref{eq:MdotMachida} by 2, 5 and 10, in order to look for a
  threshold below which migration would win over accretion. This would
  be equivalent to multiply $A$ by the corresponding number in
  \eqref{eq:ratioallmax}, hence more migration is expected. We have
  also made a simulation in which the value of $\Sigma$ in
  \eqref{eq:MdotMachida} is the actual azimuthally averaged value of
  $\Sigma$ at $a$, smaller than $\Sigma_0(a)$ when a gap
  opens. Finally, we have also prescribed an accretion rate equal to
  the radial gas flow through the disc, or the accretion rate by the
  star\,:
\begin{equation}
\dot{M}_p=\dot{M}_{\rm disc}=3\pi\nu\Sigma_0\ .
\label{eq:3pnS}
\end{equation}
  Indeed, \citet{LubowDAngelo2006} show that the planet can accrete a
  maximum of $80\%$ of what the disc's accretion rate is. For this
  reason, many authors \citep[especially in the population synthesis
    field, e.g.][]{Dittkrist-etal-2014,Bitsch-etal-2015b} consider
  that the accretion of a giant planet is limited by the accretion
  rate of the disc. We come back to this issue in the next section.}

\begin{figure}
\includegraphics[width=\linewidth]{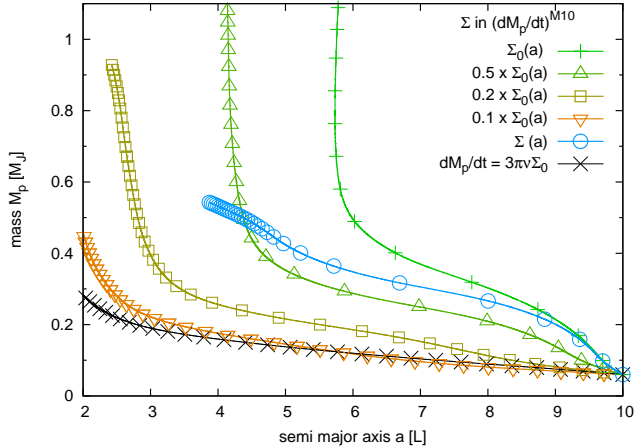}
\caption{\ac{Same as \fig{fig:M} but for $\Sigma_0=10^{-4}$ and
    various prescriptions for $\Sigma$ in \eqref{eq:MdotMachida}, as
    indicated by the key\,: $\Sigma_0(a)$ is the reference case, where
    we take the unperturbed value of the disc density at the location
    of the planet. The case $\dot{M}_p=3\pi\nu\Sigma$ is explicitely
    labelled. Again, symbols are placed every $1000\,P_L$ on the
    curves.} The threshold mass for gap opening by gravity is again
  $0.436\,M_J$.}
\label{fig:accr}
\end{figure}

\ac{The results are shown in \fig{fig:accr}. First, using the
  perturbed value of the density $\Sigma(a)$ in
  \citet{Machida-etal-2010}'s formula instead of the unperturbed value
  $\Sigma_0(a)$, does not change much until the planet reaches $\sim
  0.2\,M_J$ (compare the green curve with $+$ and the blue curve with
  $\circ$). At larger masses, the planet starts carving a gap, so
  accretion is slower with $\Sigma(a)$ than with $\Sigma_0(a)$, and
  the blue curve does not rise as fast as the green one. Consequently,
  $0.4\,M_J$ are reached after slightly more migration, but then the
  transition to type~II migration occurs\,: the circular symbols are
  very close to each other, indicating a slow migration. Therefore,
  using $\Sigma$ or $\Sigma_0$ in \eqref{eq:MdotMachida} is a key
  issue for the final mass of giant planets, but in both cases the
  accretion rate is fast enough to open a gap before being lost by
  type~I migration. Actually, this choice only matters in a phase
  where a significant gap is already opening, so it does not affect
  our answer to the question we study here.}

\ac{Second, as expected, the smaller the accretion rate, the longer
  the planet stays in type~I migration and the further it migrates. In
  particular, if the accretion rate can not exceed a tenth of
  \citet{Machida-etal-2010}'s prescription, or is limited to the gas
  flow through the disc, the planet loses $80\%$ of its semi-major
  axis before reaching the gap opening mass. In order for planets to
  have a chance of opening a gap before migrating significantly, the
  runaway accretion rate should be of the same order of magnitude as
  the one found by \citet{Machida-etal-2010}. This is consistent with
  the results of \citet{Bitsch-etal-2015b} who used an accretion rate
  limited to $80\%$ of \eqref{eq:3pnS} and found that giant planets
  observed today at $\sim 5$\,AU should be born at a few tens of AU
  and migrate a lot.}

\begin{figure}
\includegraphics[width=\linewidth]{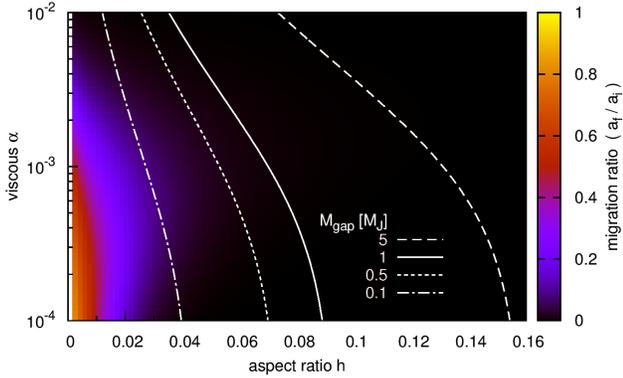}
\caption{\ac{Same as \fig{fig:migaplane}, where the planet grows at
    the rate given by \eqref{eq:3pnS} instead of the runaway regime of
    \eqref{eq:MdotMachida}. The planets still migrate in type~I
    migration following \eqref{eq:typeI} with $A=2.5$. Colour map\,:
    $a_f/a_i$ given by \eqref{eq:rationugap}. Lines\,: contours of the
    gap opening mass $M_{\rm gap}$ as given by \eqref{eq:Mgap}.}}
\label{fig:migaplanu}
\end{figure}

\ac{In fact, using $\dot{M}_p=\dot{M}_{\rm disc}$ given by
  \eqref{eq:3pnS} instead of $\dot{M}_{p,M10}$ given by
  \eqref{eq:MdotMachida}, one finds $\mathcal{G}= -
  \displaystyle\frac{A}{3\pi\alpha\,h^4}\frac{M_p}{M_*}$, hence
$$
\ln\left(\frac{a_f}{a_i}\right) =  -\frac{A}{6\pi\alpha h^4}\left[\left(\frac{M_f}{M_*}\right)^2-\left(\frac{M_i}{M_*}\right)^2\right]
$$
  and the initial to final semi-major axis ratio to reach the gap
  opening follows\,:
\begin{equation}
\ln\left(\frac{a_f}{a_i}\right) > - \frac{A}{6\pi\alpha
  h^4}\left(\frac{M_{\rm gap}}{M_*}\right)^2\ .
\label{eq:rationugap}
\end{equation}
  This is shown in \fig{fig:migaplanu}, and illustrates the fact that
  the accretion rate given by \eqref{eq:3pnS} is too low in most discs
  for the gap opening mass to be reached before type~I migration
  drives the growing planet close to its host star. Keep in mind that
  the outcome of \fig{fig:migaplanu} is strongly influenced by the
  migration rate given by $A$, which we determined here for pure
  isothermal discs with a density slope of $\alpha_\Sigma=1$. In
  reality, the inward migration can be significantly reduced by the
  entropy driven corotation torque \citep{BaruteauMasset2008AD}, or a
  shallower profile of the density profile. This may allow giant
  planet formation far away from the central star also in discs with
  higher viscosity and $h$, like in \citet{Bitsch-etal-2015b}. Still,
  this viscous accretion rate induces much more migration than the
  unlimited runaway growth \eqref{eq:MdotMachida}. Hence, in the next
  section, we address the question of the sustainability of a high
  accretion rate until a gap opens.}

  
\section{Gap opening induced by accreting giant planets}
\label{sec:suck}


\citet{Kley1999} provided a recipe for implementing accretion of gas
by a planet in 2D simulations, which has been widely used in the
literature. At every time-step, gas is removed from the cells inside
$0.45$ Hill radius of the planet and added to the planet mass, at such
a rate that $2/3$ of the gas is accreted by the planet within a given
time that can be set by a parameter (generally an orbital period). The
same applies for cells in an annulus between $0.45$ and $0.75$ Hill
radius of the planet, with an accretion time two times longer. The
linear momentum of the gas removed from the cells should also be
transferred to the planet, but we have checked that this does not
change the results. This recipe is meant to mimic an unlimited gas
collapse onto the planet, within the limits of what the disc can
provide.

We have performed an additional set of simulations in which the
accretion is performed using \citet{Kley1999}'s recipe (the mass
increase of the planet is taken from the gas disc), limited to
\citet{Machida-etal-2010}'s Eq.~(\ref{eq:MdotMachida}) in the case
where \citet{Kley1999}'s recipe would promote a too fast
accretion. The corresponding curves are the thin curves with $+$ signs
in Fig.~\ref{fig:M}. For the more massive discs, they follow closely
the thick curves with dots. This is because the mass needed to accrete
up to Jupiter's mass is already there. \ac{The gas accreted from the
  planet originates from streamlines coming close from the separatrix,
  between the circulating and the horseshoe streamlines \citep[see
    e.g. Fig.~10 of][]{Szulagyi-etal-2014}. These streamlines are fed
  from material viscously spreading from the corotation region and
  from material from the inner/outer disc. In the classical picture of
  gap opening, this material is then scattered away by the planet
  until the corotation region is depleted and a gap has opened. The
  only difference compared to our situation is now that the material
  is accreted instead of being pushed away. The horseshoe region thus
  serves as a reservoir for the planet, which feeds the accretion
  streams at the fast rate of viscous spreading across the short
  horseshoe width. As soon as all the material originating from the
  corotation region is accreted, the gap has opened, and accretion
  slows down.}

\ac{In dense discs}, the mass in the horseshoe region is of the order
of the mass of Jupiter as can be seen from Eq.~(\ref{eq:cmdmax})\,:
${\rm cmd}_{\rm max}=0.87\,M_J$ with $a=10\,L$, $\Sigma_0(L)=10^{-4}$
and $M_p=M_J$. Hence, only at the very end of the track can one see
that the accretion slows down (the $+$ signs are closer to each other
than the dots of the corresponding curve), but the planet has already
opened a gap and left the type~I migration regime\ac{. In} the case of
the lower mass discs, less mass is available in the horseshoe region
of the planet, and once it has all been accreted onto the planet, its
accretion rate drops. However, as the horseshoe region is now empty,
the planet resides in a gap. As a consequence, it does not migrate in
type~I migration, but much slower, in type~II. Therefore, in
figure~\ref{fig:M}, the tracks of these planets turn and become
vertical earlier than their unlimited accretion counterparts (for
which gas was not removed).

\begin{figure}
\includegraphics[width=\linewidth]{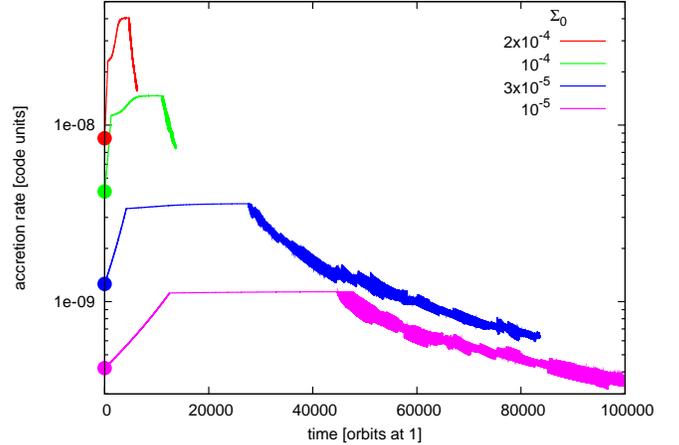}
\caption{Accretion rates of the planets corresponding to the thin
  curves with $+$ of figure~\ref{fig:M} as a function of time.}
\label{fig:Mdot}
\end{figure}

Figure~\ref{fig:Mdot} show the accretion rates of these planets as a
function of time. The initial accretion rate, marked with a dot, is
proportional to the initial surface density, as these planets all
have the same initial mass. The accretion rate then increases as $r_H$
increases, until the limiting factor is not any more the Hill radius in
Eq.~\ref{eq:MdotMachida} but the constant term. The accretion rate
then reaches a plateau, with a slight increase as the planets migrate
inwards towards regions where $\Sigma_0(r)$ is larger. At some point
though, the planets have depleted their horseshoe region, and the gas
supply limits the accretion rate, which then keeps decreasing as the
gap deepens. At this turnover point, the planet has reached a mass of
respectively $0.99,\ 0.96,\ 0.64,\ 0.36\ M_J$ in the cases where
$\Sigma_0(L)=2,\ 1,\ 0.3,\ 0.1\ \times 10^{-4}\,M_*/L^2$.

\begin{figure}
\includegraphics[width=\linewidth]{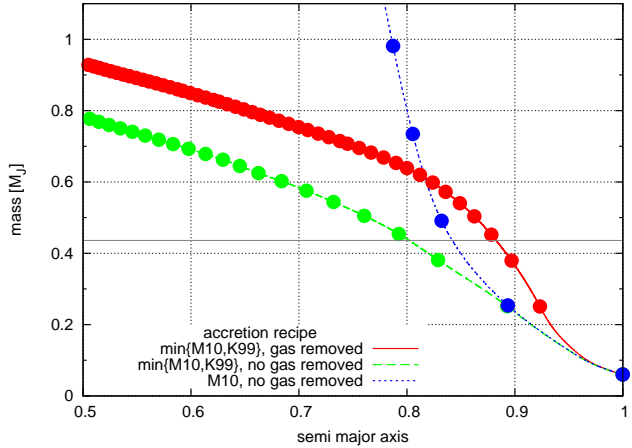}
\caption{Tracks of planets in the Mass$-$semi-major-axis plane with
  different accretion recipes. \textbf{Blue (short dashed)\,:}
  Accretion rate given by Eq.~(\ref{eq:MdotMachida}). \textbf{Red
    (solid)\,:} Accretion performed following \citet{Kley1999}'s
  recipe, but caped by Eq.(\ref{eq:MdotMachida}). \textbf{Green (long
    dashed)\,:} Same mass evolution as the red curve, but no gas is
  removed from the simulation.  \textbf{Grey\,:} The thin horizontal
  plain line marks $0.436\,M_J$, the threshold mass for gap opening by
  gravity.}
\label{fig:M-a_1}
\end{figure}

In the lightest disc simulation above, it appears that the planet
empties its horseshoe region before it reaches the gap opening
mass. This is illustrated even more clearly by an other set of
simulations, where $\Sigma_0(L)=10^{-4}$ and the planet starts at
$a=L$, so that ${\rm cmd}_{\rm max}=0.087\,M_J$ only, for a Jupiter
mass planet. Here, the resolution is $dr=0.005$ uniform,
$d\theta=0.01$, the 2D grid extends from 0.4 to 2.2 and the 1D grid
from 0.02 to 50\,; again, $h=0.05$ and $\alpha=10^{-3}$. In
figure~\ref{fig:M-a_1}, the blue short dashed curve corresponds to an
imposed mass growth given by Eq.~(\ref{eq:MdotMachida}), and no gas
removed from the disc. There is one dot every \ac{$1000\,P_L$} along the
curve. Migration slows down as the planet grows, because it opens a
gap thanks to its gravitation (and here, the cmd is negligible). The
red plain curve, however, corresponds to a case where the accretion is
performed using \citet{Kley1999}'s recipe, but limited to
\citet{Machida-etal-2010}'s rate\,; here, gas is removed from the disc
and added to the planet's mass. It is not until $900\,P_L$ that the
accretion rate drops due to lack of gas in the Hill sphere of the
planet (the first dot is almost at the same height for all the
curves), but already before this point, the planet migrates slower
than in the previous case.

From this point on, the planet migrates much slower than in the
previous case, as indicated by how close to each other the dots
are. This planet is not in type~I migration, but in type~II. Indeed,
type~II migration in this simulation is slow but inwards, as our inner
disc is accreted by the central star. However, the gas accretion rate
of the planet is so low that it barely reaches a Jupiter mass before
migrating down to $r=0.5\,L$.

\begin{figure}
\includegraphics[width=\linewidth]{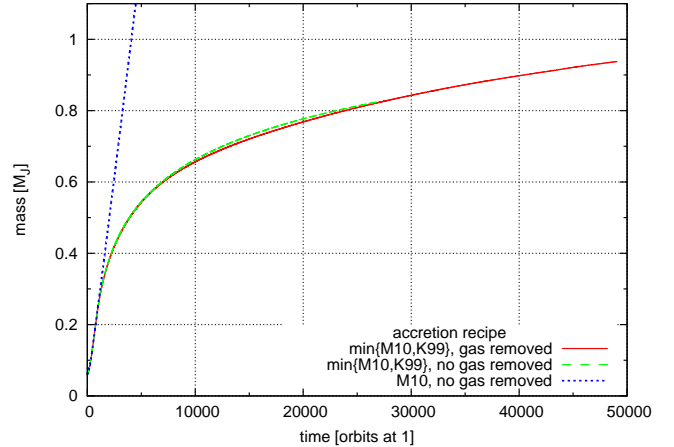}
\caption{Planetary mass as a function of time in the simulations shown
  in Fig.~\ref{fig:M-a_1}.}
\label{fig:M-t_1}
\end{figure}

Figure~\ref{fig:M-t_1} shows the mass evolution of the planet in the
cases shown in \fig{fig:M-a_1}. In the blue case, accretion is
almost linear in time. In the red case, it looks more logarithmic. In
fact, the accretion rate after $900\,P_L$ can be very well fitted by
the following expression\,:
\begin{equation}
\frac{d(M_p/M_*)}{d(t/P_L)} = 6\pi\times
10^{-8}\left(\frac{t/P_L}{1000}\right)^{-1.05} + 1.6\times
10^{-8}\exp\left(-\frac{t/P_L-900}{140}\right)\ ,
\label{eq:KM}
\end{equation}
where $t$ is the time since the beginning of the simulation. The green
curve in figures~\ref{fig:M-a_1} and~\ref{fig:M-t_1} is the track of a
planet whose mass evolution is given by \eqref{eq:MdotMachida}
before $900\,P_L$, and by \eqref{eq:KM} after, but no gas is
removed from the disc. Its mass evolution follows closely the red one,
but its migration is much faster. By the time it reaches
$M_p=0.6\,M_J$, it has migrated about twice as much as in the red case
(0.314 versus 0.177 inwards), and it reaches $a=0.5$ with as mass of
$0.78\,M_J$ versus $0.93\,M_J$ in the case where gas is removed from
the disc (red curve). This illustrates the importance of taking
material off the gas disc when a planet accretes. By emptying its
horseshoe region as it grows, a planet can save a large fraction of
the type~I migration it should suffer otherwise.

\section{Discussion}
\label{sec:discuss}

The accretion rate of giant planets is still very
uncertain. \citet{Machida-etal-2010}'s rate is probably an
overestimate, essentially because the gap opening and decrease of gas
supply is not taken into account (which would not be a concern for the
beginning of the runaway growth phase anyway), and also because of the
simplified equations of state used \citep[although this parameter
  seems to have little influence on their results\ac{, but see
    also}][]{KlahrKley2006,Uribe-etal-2013}. Shall this rate turn out
to be overestimated by an order of magnitude, \ac{the migration time-scale
  would be} too short compared to the accretion time-scale. However,
other works tend to confirm their result and the order of magnitude of
the accretion rate in the runaway regime
\citep[e.g.][]{Ayliffe-Bate-2009a,Gressel-etal-2013,Szulagyi-etal-2014}.

On the other hand, \citet{Kley1999}'s recipe in 2D simulations
probably underestimates the gas supply, because gas actually
constantly flows in the corotation region through a 3D meridional
circulation \citep{Morbidelli-etal-2014}, which is not modelled by 2D
simulations. \ac{The before mentioned simulations by
  \citet{LubowDAngelo2006} calculating the mass flow through the gap
  of an accreting planet were also performed in 2D.} Taking the
meridional circulation into account would only result in larger
accretion rates onto planets which have opened a gap. Thus, this would
not affect our conclusion that they can grow to $M_J$ at their own
pace while being in type~II migration.

Interestingly enough, integrating the first term of \eqref{eq:KM}
from $t=900\,P_L$ to infinity (the second term being negligible beyond
$t=1000\,P_L$) converges to a total of $M_{\rm accr}=3.8\,M_J$, and only
$1.12\,M_J$ in one million orbits. But one should keep in mind that
\citet{Kley1999}'s recipe is pessimistic in terms of accretion rate,
and that the numerical coefficient in \eqref{eq:KM} should be
proportional to $\Sigma_0$ so that the final mass could be
anything. In addition, in an other simulation where the same accretion
recipe was applied, we found a power index of $-0.95$ for the
accretion rate as a function of time, which diverges (still
slowly). The idea that giant planets grow roughly logarithmically in
time is attractive to explain why most extrasolar giant planets have
not reached several Jupiter masses, but gas accretion is a complex
process, and the question of the final mass of giant planets is not
the object of this paper.

\ac{We additionally performed simulations where we have taken the
  angular momentum exchange between the accreted material and the
  planet into account. It turned out that the angular momentum
  transferred from the accreted material onto the planet is of the
  order of one percent and thus negligible. This value was also found
  in the 2D studies of \citet{Duermann-Kley-2015}.  }

\section{Conclusion}
\label{sec:conclu}

In this paper, we have performed 2D, locally isothermal simulations of
\ac{embryos of} giant planets migrating in their protoplanetary disc,
while accreting gas in the runaway phase, starting at
$0.06\,M_J=20\,M_\oplus$. Our \ac{study} can be summarized as the
following points\,:

\begin{enumerate}
\item \ac{We stress that} even a fully formed Jupiter mass planet,
  once released in a disc, opens a gap in $\sim 100$ orbits. This
  remains true even if the planet migrates by more than its horseshoe
  width during this process.
\item \ac{We remind that} the transition between the fast type~I and
  the slower type~II migration regimes is not smooth in massive
  discs\,: due to the positive feedback from the coorbital mass
  deficit \citep{Masset-Papaloizou-2003}, the planet migrates faster
  than the type~I rate as it starts opening a gap, before it sharply
  transitions to the type~II regime. \ac{In very massive discs, the
    planet could have time to migrate very far or enter the type~III
    migration regime before opening its gap.}
\item \ac{We find that} the runaway growth rate given by
  \citet{Machida-etal-2010} is comparable to the type~I migration rate
  \ac{in discs with typical aspect ratios ($h\gtrsim 0.04$),} so that
  once in runaway growth, a gaseous planet can reach \ac{half} the
  mass of Jupiter without being lost into its host star by type~I
  migration. \ac{More specifically, for $\alpha\lesssim 0.005-h/30$,
    the gap opening mass $M_{\rm gap}$ can be reached in runaway
    growth while losing less than $60\%$ of the semi-major axis in
    pure type~I migration (see Fig.~\ref{fig:migaplane}).}
\item A planet in the runaway growth regime can open a gap before it
  reaches the critical mass to do so gravitationally. Indeed, as the
  planet accretes gas flowing from the separatrix around its
    horseshoe region, this region spreads and empties.
\item Independent of whether a planet has emptied its corotation
  region by accreting the gas it encompassed or by repelling the gas
  gravitationally, it is then in type~II migration, not type~I any
  more.
\item All planets start their runaway growth embedded in the disc,
  hence at this time, \citet{Machida-etal-2010}'s rate applies. Only
  once the planet's corotation region empties does $\dot{M}_p$
  decrease. The total mass available in the corotation region is
  roughly $\Sigma(a)\, a^2$ for a Jupiter mass planet, which can be of
  the order of Jupiter's mass far enough from the star or in massive
  enough discs. Hence, limiting the planetary accretion rate to the
  accretion rate through the disc towards the central star is not
  appropriate until all this available mass has been taken.
\end{enumerate}

To synthesize these points\ac{,} either a giant planet is in type~I
migration but then it is fully embedded in the disc and accretes fast
enough to open a gap before migrating too far, or its accretion is
limited by the gas supply from the disc because it has opened a deep
gap and is in the type~II migration regime. Thus, a planet in the
runaway growth is saved from type~I migration. Its subsequent type~II
migration and its final mass and position remain open questions that
we do not address in this paper.

\subsection*{Acknowledgements}
{\it Computations have been done on the 'Mesocentre SIGAMM' machine, hosted
by Observatoire de la C\^ote d'Azur. A.\,C. thanks R. Ligi for her
help with analytics and hosting at LAM. B.\,B. thanks the Knut and
Alice Wallenberg Foundation for their financial support. This work is
part of the MOJO project, ANR-13-BS05-0003-01.  We thank the two
reviewers for their remarks that contributed to significant
improvement of this paper.}

\bibliographystyle{elsarticle-harv}
\bibliography{./crida.bib}

\end{document}